\documentclass[twocolumn,letterpaper,amsmath,amssymb,floatfix,prl,superscriptaddress,showpacs]{revtex4}

\usepackage{graphicx}
\usepackage{dcolumn}
\usepackage{bm}
\usepackage{color}

\begin{document}

\title{Strong Coupling Electrostatics in the Presence of Dielectric Inhomogeneities}

\author{Y.S. Jho}
\affiliation{Materials Research Laboratory, University of
California, Santa Barbara, CA 93106, USA}
\affiliation{Dept. of Physics, Korea Advanced Institute of Science and Technology,
Yuseong-Gu, Daejeon, Korea 305-701}

\author{M. Kandu\v c}
\affiliation{Dept. of Theoretical Physics, J. Stefan Institute,
SI--1000 Ljubljana, Slovenia}

\author{A. Naji}
\affiliation{Materials Research Laboratory, University of
California, Santa Barbara, CA 93106, USA}
\affiliation{Dept. of Chemistry and Biochemistry,
University of California, Santa Barbara, CA 93106, USA}

\author{R. Podgornik}
\affiliation{Dept. of Theoretical Physics, J. Stefan Institute,
SI--1000 Ljubljana, Slovenia}
\affiliation{Dept. of Physics, Faculty of Mathematics and Physics, University
of Ljubljana, SI--1000 Ljubljana, Slovenia}

\author{M.W. Kim}
\affiliation{Materials Research Laboratory, University of
California, Santa Barbara, CA 93106, USA}
\affiliation{Dept. of Physics, Korea Advanced Institute of Science and Technology,
Yuseong-Gu, Daejeon, Korea 305-701}

\author{P.A. Pincus}
\affiliation{Materials Research Laboratory, University of
California, Santa Barbara, CA 93106, USA}
\affiliation{Dept. of Physics, Korea Advanced Institute of Science and Technology,
Yuseong-Gu, Daejeon, Korea 305-701}


\begin{abstract}
We study the  strong-coupling (SC) interaction between
two like-charged membranes of finite thickness embedded in a medium of higher
dielectric constant. A generalized SC theory is applied along with
extensive Monte-Carlo simulations to study the image charge effects induced
by multiple dielectric discontinuities in this system. These
effects lead to strong counterion crowding in the central region
of the inter-surface space upon increasing the solvent/membrane
dielectric mismatch and change the membrane interactions from attractive to repulsive
at small separations. These features agree quantitatively with the SC theory at 
elevated couplings or dielectric mismatch where 
the correlation hole around counterions is larger than
the thickness of the central counterion layer.
\end{abstract}


\maketitle


Biological macromolecules such as DNA, lipid membranes and
proteins are highly charged in water. Electrostatic
interactions play a key role in determining structure, phase
behavior and specific functioning of these macroions in aqueous biological
media \cite{Andelman}. One particular trait of these systems
is that their behavior is dominated, to a large extent, by
neutralizing counterions that surround them in a diffuse ionic
cloud. When present at higher valencies, these counterions are
known to generate strong electrostatic attractions between
like-charged macroions as observed in numerous experiments and
simulations \cite{Netz-review}. These observations stand in stark
contrast with the  traditional mean-field or Poisson-Boltzmann
(PB) theories, which predict purely repulsive forces
\cite{Andelman}.

Consequently, there have been a number of attempts to assess
corrections to the PB theory using, {\em e.g.}, correlated density
fluctuations around the mean-field distribution or additional
non-electrostatic interactions \cite{refs:fluctuations,andelman:non-ele}.
An alternative approach has been developed recently \cite{Rouzina,shklovskii,Levin,Netz,Netz-review,Netz-review2}, 
which leads to the so-called strong-coupling (SC) theory; it is known to
become exact in the limit of high macroion charge, large
counterion valency, low medium dielectric constant or low temperature,
where the PB theory breaks down. 

While the strong-coupling phenomena are well-understood in the
framework of the SC theory, it still remains a challenge to predict
the behavior of realistic biophysical systems, which, among other
things, exhibit highly inhomogeneous dielectric structure. The
large difference in static dielectric constants of water ($\varepsilon
\simeq 80$), being the most common solvent, and the non-polar
moieties ($\varepsilon \simeq 2-5$), comprising the molecular
interiors of proteins, lipid membranes and DNA, leads to
substantial differences in interactions between charges in most
common biological environments. Moreover, the presence of a
combination of both  aqueous and hydrocarbon regions in the
immediate surrounding (as, {\em e.g.}, in the case of two or more
interacting lipid membranes) leads to a more complex pattern of
{\em image charges}  that are induced by multiple
dielectric discontinuities in the system.

In this Letter, we consider a system of two like-charged membranes
of finite thickness in a medium of higher
dielectric constant and determine the counterion distribution and
the interaction between the membranes by means of
both Monte-Carlo simulations and a generalized SC theory. Recent
studies show that image interactions in highly charged systems
could result in remarkable effects such as in the surface
adsorption of flexible polyelectrolytes \cite{joanny} and the
charge inversion of macroions \cite{shklovskii-image}. 
In the slab geometry, the  dielectric discontinuity effects have been
studied in both weak-coupling
\cite{Netz-review2,bratko-kjellander,menes2000,Andre2002,Ha2005,Rudi2007}
and strong-coupling regimes
\cite{Netz-review2,Andre2002,Rudi2007,joys2007}. These works however 
deal with one or two semi-infinite slabs and do not consider
 the finite thickness of the membranes or the multitude
 of images produced in the two-slab system. 
 While accounting only for the 
 first-order images serves as an accurate approximation 
 at low couplings \cite{Netz-review2,bratko-kjellander,Andre2002}, 
 the same approximation as we show breaks down on a qualitative 
 level at elevated couplings, where the higher-order induced images play an essential role.
 For vanishing membrane thickness (no dielectric
discontinuity), the long-range SC attraction mediated by
counterions is dominant and leads to a tightly bound state between
two like-charged membranes \cite{Netz,Netz-review2}. 
 This well-known picture changes as the
membrane thickness and the solvent/membrane dielectric mismatch
increase, leading to enhanced repulsive (and even change from
attractive to repulsive) interaction between membranes at small
separations. This SC repulsion depends strongly on the dielectric
mismatch contrary to the (repulsive) PB interaction which is not
affected by the image charges.

\begin{figure}[t]
\includegraphics[width=5.2cm]{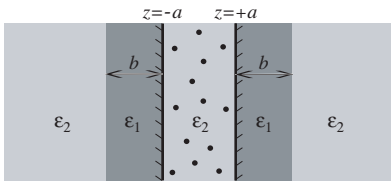}
\caption{Schematic view of two charged membranes and counterions in between. Here   
$2a$ is the closest-approach distance (leaving
out the counterion diameter from the actual distance) and counterion excluded-volume repulsions are neglected.
}
\label{fig:schematic}
\end{figure}

Let us consider two charged parallel membranes of thickness $b$
and dielectric constant $\varepsilon_1$ with their inner surfaces being
located at $z=\pm a$ and bearing uniform charge density $-\sigma e $
(Fig. \ref{fig:schematic}). The system is
immersed in a solution of dielectric constant $\varepsilon_2$
containing $+q$-valent counterions.
Apart from the direct contribution from
the charged surfaces, the electrostatic potential experienced by
each counterion also involves contributions from image charges
induced at each solvent/membrane  interface.
We shall consider the contribution from all image charges in the
present geometry.
To this end, we have developed a new numerical algorithm that enables one
to compute efficiently the electrostatic interactions in the presence of multiple
dielectric discontinuities  \cite{joys2008}.
Assuming two-dimensional
lateral periodicity (in the plane of the membranes),
we combine the image charge method with the
MMM2D summation technique to obtain fast-converging
series for evaluating Coulombic interactions \cite{Arnold2002,joys2007}.

The state of the system may be described in terms of
the following dimensionless parameters: the {\em dielectric jump
parameter} $\Delta =
(\varepsilon_2-\varepsilon_1)/(\varepsilon_2+\varepsilon_1)$, the rescaled
half-distance $\tilde a =a/\mu$, the rescaled thickness $\tilde
b=b/\mu$, and the {\em electrostatic coupling parameter}
$\Xi=q^2\ell_{\mathrm{B}}/\mu$, where  $\mu = 1/(2\pi q
\ell_{\mathrm{B}} \sigma)$  is the so-called Gouy-Chapman length
and $\ell_{\mathrm{B}}=e^2/(4\pi \varepsilon_2 \varepsilon_0
k_{\mathrm{B}}T)$ is the Bjerrum length \cite{Andelman,Netz-review}. 
The coupling parameter $\Xi$ determines the
strength of electrostatic correlations: for dielectrically homogeneous systems,
the SC behavior dominates typically for $\Xi > 10$, while the PB description
is found to be valid at small couplings $\Xi<1$
\cite{Netz-review,Netz,Netz-review2}.

For a typical membrane with $\sigma e\simeq
1\frac{e}{\mathrm{nm}^2}$ in water ($\varepsilon_2\simeq  80$ at room temperature), we have
$\Xi\simeq 3, 25, 85$ and 200 by choosing counterion valency as
$q=1,\ldots, 4$, respectively. Higher couplings
may be obtained by using less dielectric solvents ({\em
e.g.}, mixtures containing methanol, $\varepsilon_2\simeq 33$) since $\Xi\sim 1/\varepsilon_2^2$.
Here we shall focus mainly on the SC regime with large $\Xi$ and
vary $\Delta$ (at fixed $\varepsilon_2 = 80$)
in the range $\Delta=0$ (no discontinuity, $\varepsilon_1 = 80$) to
$\Delta=0.95$ (water/hydrocarbon interface, $\varepsilon_1 = 2$).

\begin{figure*}[t]
\includegraphics[width=16.cm]{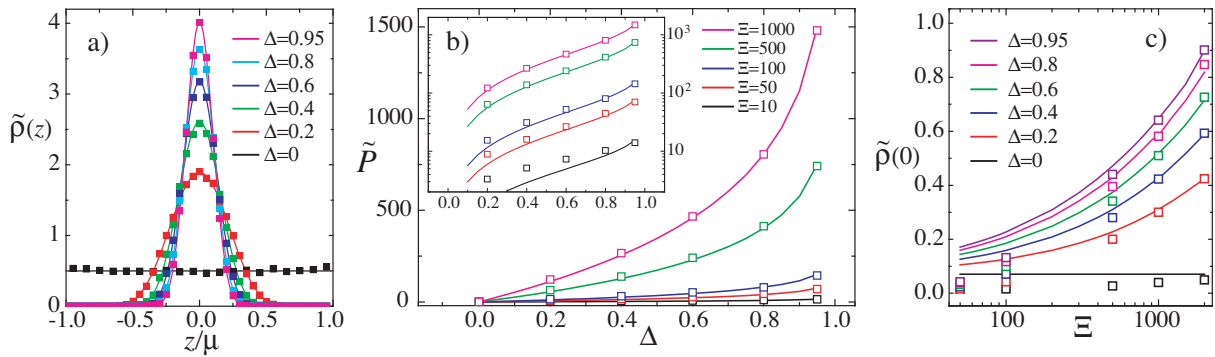}
\caption{a) Rescaled counterion density profile
between two like-charged dielectric membranes for $\Xi=100$,
$b/\mu=100$, $ a/\mu=1$ and as  the dielectric mismatch $\Delta$ is increased (from bottom to top at $z=0$). 
b) Rescaled inter-membrane pressure as a function of
$\Delta$ for $b/\mu=100$, $ a/\mu=1$ and as 
$\Xi$ is increased (bottom to top); inset shows the details in
the logarithmic scale.
c) The mid-plane ($z=0$) counterion density, $\tilde \rho(0)$, 
as a function of the coupling parameter $\Xi$ for $a/\mu=7$,  
$b/\mu=100$ and as $\Delta$ is increased (bottom to top). 
In a)-c) lines are the SC predictions (see the text)
and symbols are simulation data.  }
\label{fig:fig2}
\end{figure*}

In Fig. \ref{fig:fig2}a, we show simulated
counterion density profiles between two membranes of
thickness $b/\mu=100$ and coupling parameter $\Xi=100$ as
$\Delta$ varies  (symbols).  As seen, counterions are strongly depleted from the vicinity
of the charged surfaces and accumulate around the mid-plane 
as the dielectric mismatch is increased--a trend
observed also in studies that include only the first induced images \cite{Netz-review2,bratko-kjellander}.
This behavior is intimately connected with the electrostatic
correlations, {\em i.e.}, it will be absent in the PB
limit $\Xi\rightarrow 0$ (see Fig. \ref{fig:fig2}c). Intuitively, for
$\varepsilon_2>\varepsilon_1$
counterions have the same sign as their images and the
counterion-image repulsion provides the mechanism for the
foregoing observation in the SC regime.

In general, the origin of SC phenomena goes back to the fact that at large
couplings, $\Xi\gg 1$, counterions strongly repel each other and tend to form a
highly correlated quasi-2D layer close to a charged surface. Thus,  individual ions become
increasingly ``isolated'' as they are surrounded by a large
correlation hole of size $a_\bot^2 \sim q/\sigma$ (or $a_\bot/\mu
\sim \sqrt{\Xi}$) as follows from the local electroneutrality
condition \cite{Netz-review2}. As a result, the single-particle
interaction between individual counterions
(including their own images) and the charged membranes, $u(\tilde z)$, is expected to
dominate and lead to a barometric number density profile, $ \rho_{\mathrm{SC}}( z)$,
for counterions; in rescaled units
$\tilde \rho_{\mathrm{SC}}(\tilde z)  \equiv \frac{\rho_{\mathrm{SC}}(z)}{4\pi\ell_{\mathrm{B}}\sigma^2} = A(\tilde a)\, e^{-u(\tilde z)}$,
where $\tilde z = z/\mu$ and $1/A(\tilde a)=\int_{-\tilde a}^{\tilde a}{\mathrm{d}}\tilde z\,\tilde
\rho_{\mathrm{SC}}(\tilde z)$.
For $\Delta=0$, this result indeed
follows as an exact limiting ($\Xi\rightarrow \infty$) result  from
a systematic $1/\Xi$-expansion \cite{Netz-review,Netz},  giving
a {\em uniform} profile $\tilde \rho_{\mathrm{SC}}(\tilde z) =
\frac{1}{2\tilde a}$ (with $u(\tilde z)=0$ and $A(\tilde a)=\frac{1}{2\tilde
a}$) in agreement with our simulations (horizontal line in Fig.
\ref{fig:fig2}a). Higher-order terms in general involve
corrections due to counterion-counterion interactions at finite $\Xi$ \cite{Netz}.

For an inhomogeneous system with
$\Delta>0$, we generalize the $1/\Xi$-expansion method by taking into account
the presence of all four dielectric boundaries (Fig. \ref{fig:schematic}) and
obtain the leading-order ($\Xi\rightarrow \infty$) SC density profile  as
\begin{equation}
   \tilde \rho_{\mathrm{SC}}(\tilde{z}) = A(\tilde a) ~e^{ -\Xi \int_{0}^{\infty}\!{\mathrm{d}} Q\,
   \frac{\cosh 2Q\tilde{z}}
   {\Delta_Q^{-1}e^{2Q\tilde a}-\Delta_Q  \,e^{-2Q\tilde a}}},
   \label{eq:SC_density}
\end{equation}
where $\Delta_Q  =
\Delta\frac{1-\exp(-2Q\tilde b)}{1-\Delta^2\exp(-2Q\tilde b)}$.
This result is plotted in Fig. \ref{fig:fig2}a (solid
lines) and is in excellent agreement with our data
(symbols) for all dielectric jump parameters at $\Xi=100$.  Qualitatively, the
bellshaped profiles 
may be approximated for large $b$ by the first-image expression
\begin{equation}
  \tilde \rho_{\mathrm{SC}}(\tilde{z}) \simeq A(\tilde a)\,e^{-\frac{\Delta\Xi}{2}\frac{\tilde a}{\tilde a^2-\tilde z^2}}.
  \label{eq:approx_density}
\end{equation}

The SC pressure acting on each membrane may be calculated
from $\tilde{P}_{\mathrm{SC}}\equiv
P_{\mathrm{SC}}/(2\pi\ell_{\mathrm{B}}\sigma^2 k_{\mathrm{B}}T) =-\partial
\tilde{\mathcal F}_{\mathrm{SC}}/\partial \tilde{a}$ by evaluating
the SC free energy $\tilde{\mathcal F}_{\mathrm{SC}}$
(per $k_{\mathrm{B}}T$ and number of counterions)
using standard SC methods \cite{Netz-review2,Rudi2007}; hence
\begin{equation}
   \tilde{\mathcal F}_{\mathrm{SC}} = \tilde{a} - \ln \int_{0}^{\tilde{a}}\!{\mathrm{d}}\tilde{z}\,\,
   e^{ -\Xi \int_{0}^{\infty}{\mathrm{d}} Q
   \frac{\cosh 2Q\tilde{z}+\Delta_Q \,e^{-2Q\tilde{a}}}
   {\Delta_Q^{-1}e^{2Q\tilde a}-\Delta_Q  \,e^{-2Q\tilde a}}}.
   \label{eq:SC-free}
\end{equation}
The first term is the SC energetic attraction induced by single counterions
between the two surfaces when there are
no images ($\Delta=0$) \cite{Netz,Netz-review2}. The
second term includes the counterion confinement entropy
as well as the image charge contributions.
For $\Delta=0$, this second term reduces to the entropic $\ln
\tilde a$ term as expected and the SC pressure follows as
$\tilde{P}_{\mathrm{SC}} = -1+1/\tilde a$, featuring
repulsion at small separation, $a<a_\ast$, and attraction at large separation, $a>a_\ast$,
with $a_\ast/\mu=1$ being the stable equilibrium (zero-pressure) half-distance.
Upon increasing $\Delta$, the inter-membrane pressure
 becomes increasingly more repulsive at small separations
due to the counterion-image repulsions; it may be increased by up to an order of magnitude
as shown by the data in Fig. \ref{fig:fig2}b (symbols).
This behavior is captured quantitatively by the SC pressure, $\tilde{P}_{\mathrm{SC}}$ (solid lines,
obtained from Eq. (\ref{eq:SC-free})) for high enough couplings at all
values of the dielectric jump.

So far, we have considered only small membrane separations $a\sim \mu$,  
where good agreement with the SC
predictions is achieved for $\Xi\sim 100$ and larger couplings. Formally,
the SC  theory is exact for all separations
 when  $\Xi\rightarrow \infty$. But for finite $\Xi$ as is the
case in simulations, this limiting single-particle description
is expected to perform poorly at large
separations, where the sub-leading counterion-counterion
interactions  play a more significant role \cite{Netz}. For
$\Delta=0$, the SC validity regime is identified as $a \ll a_\bot$
or $\tilde a\ll \sqrt{\Xi}$ \cite{Netz-review,Netz-review2}, {\em
i.e.}, when the surface separation is smaller than the counterion
spacing, $a_\bot$, defined above. In the presence of images, counterions are
crowded in a thin layer of thickness (or the full width at half
maximum) $\delta_z<a$ around the mid-plane (Fig.
\ref{fig:fig2}a). From Eq. (\ref{eq:approx_density}) the thickness $\delta_z$ may be estimated as
$\delta_z = a[1-(1+2a\ln 2/\Xi\Delta)^{-1}]^{1/2}$. The SC theory
is expected to hold when $a_\bot$ is larger than
$\delta_z$, {\em i.e.}, when counterions form a quasi-2D layer.
Hence, we find the generalized criterion $\delta_z \ll a_\bot$ or
(in rescaled units)
\begin{equation}
\tilde a^3 \ll \tilde a\,\Xi + \Delta\,\Xi^2.
\label{eq:criterion}
\end{equation}
Thus, for the SC theory to be valid at larger $a$,
a larger coupling parameter or dielectric mismatch is needed.
Moreover, this predicts that the SC theory remains valid for
a {\em larger} range of separations when $\Delta$ is nonzero.

\begin{figure*}[t]
\includegraphics[width=16.cm]{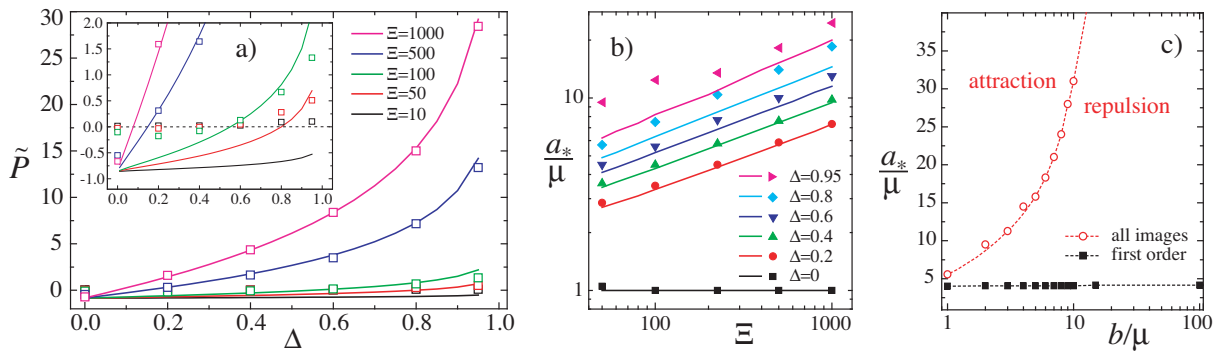}
\caption{a) Same as Fig. \ref{fig:fig2}b but for  $a/\mu=7$. Inset shows the details for small pressures. b) Simulated equilibrium half-distance, $a_\ast$, 
as a function of $\Xi$ for $b/\mu=2$ and as $\Delta$ is increased. Lines are SC predictions
(from minimization of Eq. (\ref{eq:SC-free})). 
c) Simulation results for $a_\ast$ as a function of $b$ with  all images (circles) and 
the first-order images only (squares) for $\Xi=50$ and $\Delta=0.95$ (the latter shows a very weak
increase with $b$); here
 lines are guides to the eye. The surfaces
attract for $a>a_\ast$ and repel for $a<a_\ast$. }
\label{fig:fig3}  
\end{figure*}

In Fig. \ref{fig:fig2}c we show the mid-plane density
$\tilde \rho(0)$ as a function of $\Xi$ for a larger
half-distance $a/\mu =7$. The SC theory (solid lines) is no more valid
for $\Xi\lesssim 100$, largely overestimating the
simulated density (symbols). This reflects the fact that
counterion-counterion repulsions absent in the leading-order
theory are not negligible and tend to drive the counterions away
from the crowded mid-plane toward the surfaces.
The same trend also transpires in the
pressure data for $a/\mu =7$ (Fig. \ref{fig:fig3}a), where good agreement is obtained only for $\Xi\gtrsim 500$.

The theoretical and numerical results are generally found to be in better agreement
for larger $\Delta$ as expected from Eq. (\ref{eq:criterion}); see Figs. \ref{fig:fig2}b (inset),
\ref{fig:fig2}c and \ref{fig:fig3}a (inset).
The change from attractive to repulsive interaction upon increasing $\Delta$
is more clearly seen in Fig. \ref{fig:fig3}a.

In the presence of dielectric inhomogeneities not only the
strength of the inter-membrane repulsion becomes larger but also
its range increases, pushing the attraction regime and the
equilibrium separation, $a_\ast$, to larger distances. When
$\Delta=0$, as noted above, $\tilde a_\ast=a_\ast/\mu=1$ in
agreement with our simulations in Fig. \ref{fig:fig3}b.
For $\Delta>0$, $\tilde a_\ast$ shows a monotonic increase with
the coupling parameter. While the qualitative trend is reproduced
by the SC theory, the quantitative agreement between simulated
$\tilde a_\ast$ (symbols) and the theoretical predictions (solid lines)
is obtained only for small to intermediate $\Delta$. In general,
the SC theory underestimates the $\tilde a_\ast$ values (by
$\lesssim 20$\% in Fig.  \ref{fig:fig3}b). This
discrepancy reflects the interplay between counterion-image
repulsions (included in both theory and simulations) and
counterion-counterion repulsions (included only in the
simulations): as $\Delta$ increases, the former grows pushing
$a_\ast$ to larger values beyond the regime of validity of the SC
theory (Eq. (\ref{eq:criterion})), where the latter effects need
to be accounted for as well. A similar trend is found when
membrane thickness is increased.

Note that while the weak-coupling results are less sensitive to the dielectric jump effects, 
a remarkable dependence on $\Delta$ is found in the SC regime 
(Figs. \ref{fig:fig2}b-c and \ref{fig:fig3}a). Since the SC interactions are long-ranged 
(within the regime defined by Eq. (\ref{eq:criterion})), 
a larger  number of 
induced images are needed to be taken into account in the SC regime. The first-order-image approximation \cite{Netz-review2,bratko-kjellander} fails qualitatively at high couplings and particularly for thick membranes 
as it always predicts a strong attraction and a remarkably different (smaller) $a_\ast$ as shown in Fig. \ref{fig:fig3}c.

In conclusion, we have shown that the generalized SC theory can
explain the effects of dielectric inhomogeneity in the two-slab
system at elevated couplings (specified by an extended SC
criterion), where counterions are
strongly depleted from the interfacial regions leading 
to enhanced repulsive inter-membrane interactions. 
Thus in contrast to weakly coupled systems, the electrostatics of highly coupled systems can be affected significantly by the image charges.
As an experimentally measurable effect, the SC images 
may lead to swelling of the multilamellar lipid 
arrays upon changing the solvent dielectric constant. In reality, 
however, one deals with additional factors such as smooth dielectric 
profiles and discrete surface charges \cite{Andre2002,Ha2005,joys2007}, which offer interesting problems for a more detailed study in the future.

Y.S.J., P.A.P. and M.W.K. acknowledge funds from the National Science Foundation
(Grants DMR-0503347, DMR-0710521) and MRSEC NSF 
DMR-0520415. Y.S.J. and M.W.K. have been supported by the KISTEP
(Grant I-03-064) and the Korea Health 21 R\&D Project, and M.K. and R.P.
by the Agency for Research and Development of Slovenia (Grants P1-0055(C), Z1-7171, L2-7080).


\end{document}